\documentclass[a4paper,12pt]{article}

\newcommand{\sect}[1]{\setcounter{equation}{0}\section{#1}}

\begin{document}
\topmargin 0pt
\oddsidemargin 0mm

\renewcommand{\thefootnote}{\fnsymbol{footnote}}
\begin{titlepage}
\begin{flushright}
OU-HET 388\\
hep-th/0105214
\end{flushright}

\vspace{5mm}
\begin{center}
{\Large \bf Holography and  
Brane Cosmology in Domain Wall Backgrounds}
\vspace{12mm}

{\large
Rong-Gen Cai$^{1,2}$\footnote{email address: cai@het.phys.sci.osaka-u.ac.jp}
and  Yuan-Zhong Zhang$^{3,1}$\footnote{email address: yzhang@itp.ac.cn}} \\
\vspace{8mm}
{\em $^1$ Institute of Theoretical Physics, Chinese Academy of Sciences,
  \\
    P.O. Box 2735, Beijing 100080, China \\ 
 $^2 $ Department of Physics, Osaka University,
Toyonaka, Osaka 560-0043, Japan \\
 $^3 $ CCAST (World Lab), P.O. Box 8730, Beijing 100080, China }
\end{center}
\vspace{5mm}
\centerline{{\bf{Abstract}}}
\vspace{5mm}
We consider a class of domain-wall black hole solutions in the dilaton gravity
with a Liouville-type dilaton potential. Using the surface counterterm approach
we calculate the stress-energy tensor of quantum field theory (QFT) 
corresponding to the domain-wall black hole in the domain-wall/QFT
correspondence. A brane universe is investigated in the domain-wall black hole 
background. When the tension term of the brane is equal to  the surface
counterterm, we find that the equation of motion of the brane can be mapped to
the standard form of FRW equations, but with a varying gravitational constant 
on the brane. A Cardy-Verlinde-like formula is  found, which relates the 
entropy density of the QFT to its energy density. At the moment when the brane 
crosses the black hole horizon of the background, the Cardy-Verlinde-like 
formula coincides with the Friedmann equation of the brane universe, and the 
Hubble entropy bound is  saturated by the entropy of domain-wall black holes.

\end{titlepage}

\newpage
\renewcommand{\thefootnote}{\arabic{footnote}}
\setcounter{footnote}{0}
\setcounter{page}{2}

\newpage
\sect{Introduction}

Holographic principle is perhaps  one of fundamental principles of nature,
which relates a theory with gravity in $D$ dimensions to a theory in  
$(D-1)$  dimensions  without gravity \cite{Hooft,Suss}. The AdS/CFT 
correspondence \cite{Mald,GKP,Witten1} is a beautiful example for the 
realization of the holographic principle.

The brane world scenario of the Randall and Sundrum model \cite{RS} has been
a new arena to further understand  the holographic feature of gravity 
(see, for example, \cite{Carl,Pere} and references therein). 
In Ref.~\cite{Gubser} it has been shown that a radiation-dominated flat ($k=0$)
Friedmann-Robertson-Walker (FRW) cosmology emerges as the induced metric
on a codimension one hypersurface of constant extrinsic curvature in the
background of a five-dimensional AdS Schwarzschild solution with a Ricci
flat horizon. The radiation can
be interpreted as the conformal field theory (CFT) corresponding to
the AdS Schwarzschild black hole in the AdS/CFT correspondence. 
This holographic picture has been further studied recently by Savonije and 
Verlinde \cite{Savo}, based on an observation made by Verlinde \cite{Verl}
on the relation between the Cardy formula of CFTs and the Friedmann equation
 of a closed FRW universe.  By considering a brane universe in the background 
of AdS Schwarzschild black holes in  arbitrary dimensions, except that the
 induced geometry of the brane is exactly given by that of a standard 
radiation-dominated closed FRW universe, Savonije and  Verlinde observed that 
when the brane crosses the horizon of background geometry, the entropy and 
temperature of the universe can be simply expressed in  terms of the Hubble 
constant and its time derivative; 
the entropy formula of CFTs in arbitrary dimensions, called the 
Cardy-Verlinde formula \cite{Verl}, coincides with the Friedmann equation of
the brane universe; and
the Hubble entropy bound is just equal to the entropy of the AdS Schwarzschild 
black holes\footnote{These observations  have been extended to the cases of
charged AdS black holes in \cite{BM,CMO}, and topological black holes 
in \cite{DY}. For other related discussions on the Cardy-Verlinde formula 
see \cite{Klem}-\cite{Youm}.}.

It would be of interest to see to what extent these observations on the
holographic properties  are universally valid. For this purpose, in this 
paper we investigate these holographic connections in the  case of dilaton 
domain-wall black holes as the background, replacing the AdS Schwarzschild 
black holes. In this case, according to the domain-wall/QFT 
correspondence \cite{BST,Behr,CO}, corresponding to the domain-wall 
black hole is a general quantum field theory (QFT), rather than a CFT. 
Using the method of surface counterterm, in section II  we first calculate the 
stress-energy tensor of the QFT and discuss the thermodynamics of the 
domain-wall black holes. The  brane cosmology in the background of the dilaton 
domain-wall black holes and holographic properties of the brane universe are 
discussed in section III. Our results are summarized in section IV with a 
brief discussion.


\sect{Domain-wall black holes }

We start with the action of an $(n+2)$-dimensional dilaton gravity with a 
Liouville-type dilaton potential,
\begin{equation}
\label{2eq1}
S =\frac{1}{16\pi G_{n+2}}\int d^{n+2}x\sqrt{-g}\left ( R-
   \frac{1}{2}(\partial \phi)^2 +V_0 e^{-a\phi} \right),
\end{equation}
where $G_{n+2}$ is the gravitational constant in $(n+2)$ dimensions, $V_0$ and
$a$ are assumed to be two positive constants (when $a$ is a negative constant, 
it can be changed to a positive one by replacing $\phi$ by $ -\phi$).

Such an action (\ref{2eq1}) naturally arises as 
a consistent truncation of  various supergravities. For example,  
in the decoupling limit of D$p$-branes in type II supergravity, one
can do a consistent sphere reduction in the dual frame, resulting in an 
effective action like Eq.~(\ref{2eq1}) with \cite{CO}
\begin{equation}
\label{2eq2}
V_0 = \frac{1}{2}(9-p)(7-p) {\rm \bf N}^{-2\lambda/p}, 
    \ \  a= \frac{\sqrt{2}(p-3)}
   {\sqrt{p(9-p)}}, \ \ \ \lambda = \frac{2(p-3)}{7-p},
\end{equation}
where $p=n$ and  ${\rm \bf N}$ is the number of D$p$-branes. 
In this case, when $p=3$, the dilaton potential becomes a constant, and  
one then has a five-dimensional AdS Schwarzschild black hole solution in the 
action (\ref{2eq1}) with a constant dilaton. This is consistent with the 
fact that the bulk geometry of D3-branes is a five-dimensional AdS 
Schwarzschild solution (times a 5-sphere) in the decoupling limit.

In a general case, one may consider the following Lagrangian in 
$D$ dimensions \cite{Cvetic}
\begin{equation}
\label{2eq3}
{\cal L}= R -\frac{1}{2}(\partial \phi_1)^2 -\frac{1}{2(D-n-2)!}e^{a_1\phi_1}
     F^2_{(D-n-2)},
\end{equation}
where $F_{(D-n-2)}$ is a $(D-n-2)$-form field strength and 
\begin{equation}
\label{2eq4}
a^2_1 =\frac{4}{N_1}-\frac{2(n+1)(D-n-3)}{D-2}.
\end{equation}
 $N_1$ is an integer in supergravity theories. The case $N_1=1$ can arise
for all forms in supergravity. In particular, all the field strengths have 
$N_1=1$ in $D=10$ and $D=11$ supergravities. The case $N_1=2$ can appear
for 2-forms in $D<9$, and 3-forms in $D<6$ in the non-maximal supergravities.
The case $N_1=3$ can arise for 2-forms in $D \le 5$ and $N_1=4$ for 2-forms in
$D \le 4$ \cite{Cvetic}. After a consistent sphere reduction, one can 
obtain an effective action like (\ref{2eq1}) with \cite{Cvetic}
\begin{eqnarray}
\label{2eq5}
&& V_0= \frac{2(D-n-3)^2b^2}{\triangle}, \ \ 
     \triangle =\frac{4(D-n-3)}{2(D-n-2)
     -(D-n-3)N_1}, \nonumber \\
&& a^2= \frac{2(n+1)}{n}-\frac{4(D-n-3)}{2(D-n-2)-
      (D-n-3)N_1},
\end{eqnarray}
where $b^2$ is a parameter.

There exists a class of domain wall solutions in the action (\ref{2eq1}), 
and the localization of gravity on the domain wall has been investigated
in some detail in \cite{Cvetic} (see also \cite{Youm1}). For our purposes, 
however, we need to look for solutions with black hole horizon.
Solving the equations of motion of the action (\ref{2eq1}), we find a set of 
solutions as follows, 
\begin{eqnarray}
\label{2eq6}
&& ds^2 = -f(r) dt^2 +f(r)^{-1}dr^2 +R^2(r) dx_n^2, \nonumber \\
&& R(r) =r^N,  \nonumber \\
&& \phi(r) =\phi_0 +\sqrt{2nN(1-N)}\ln \ r, \nonumber\\
&& f(r)= \frac{V_0e^{-a\phi_0}r^{2N}}{nN(N(n+2)-1)} -\frac{m r^{1-nN}}
      {\sqrt{2nN(1-N)}},
\end{eqnarray}
where $dx_n^2$ stands for the line element for a $n$-dimensional Ricci flat
space, $\phi_0$ and $m$ are two integration constants and the parameter $N$ 
obeys
\begin{equation}
\label{2eq7}
a =\frac{\sqrt{2nN(1-N)}}{nN}.
\end{equation}
Let us first discuss some special cases of the solution (\ref{2eq6}). 
When $m=0$,
the solution (\ref{2eq6}) can be transformed to the form of domain-wall 
solution in \cite{Cvetic} by dropping the "$1$" of the harmonic function $H$
there. When $m \ne 0$, the solution (\ref{2eq6}) also approaches
the domain-wall solution as $r\to \infty$ if $1-nN <2N$. When $N=1$, by 
redefining the integration constant $m$, we can see that in this case the 
solution is just the AdS Schwarzschild black hole solution with a Ricci flat
horizon, and the constant $m$ is proportional to the mass of black 
hole. In a general case where $m>0$ and $1-nN <2N$, the 
solution (\ref{2eq6}) has a black hole horizon $r_+$ satisfying the equation
$f(r_+)=0$. In this sense, we call the solution (\ref{2eq6})  domain-wall 
black hole\footnote{This term comes from Ref.~\cite{CLLP}, in 
which a set of
charged domain-wall black hole solutions with $N=1/2$ has been found. 
Such solutions have been called black plane solutions in earlier 
references~\cite{CZ}, there some charged black plane solutions have been 
obtained in four dimensions. Similar solutions have been found independently 
in \cite{Chamb}.}. When $m<0$, the singularity at $r=0$ in the 
solution (\ref{2eq6}) becomes naked. Therefore, we will not further consider 
this case.

In the AdS/CFT correspondence, the thermodynamics of AdS Schwarzschild black 
hole in the high temperature limit can be identified with that of boundary
CFTs \cite{Witten2}. Naturally we can identify the thermodynamics of 
domain-wall black holes with that of QFTs residing on the domain 
walls~\cite{CO}. In the traditional Euclidean path integral approach to
black hole thermodynamics, one usually uses the background subtraction
procedure, in which in order to get a finite Euclidean action of 
black hole and a finite quasilocal stress-energy tensor of gravitational 
field~\cite{Brow}, one has to choose a suitable reference background. 
Such a procedure
causes the resulting physical quantities to depend on the choice of reference
background. Furthermore, sometimes one may encounter situations in which 
there are no appropriate reference backgrounds. In  asymptotically AdS
spacetimes, this difficulty has been solved by adding suitable surface 
counterterms to the gravitational action~\cite{Bala}. In this way, one can
obtain a well-defined boundary stress-energy tensor and a finite Euclidean
action for the asymptotically AdS spacetimes.

Now we calculate the stress-energy tensor of the QFT corresponding to the
bulk solution (\ref{2eq6}) using the surface counterterm approach. 
In \cite{CO} (see also \cite{Nojiri}) 
we have already shown that for a class of solutions like 
(\ref{2eq6}) one can also obtain a well-defined  boundary stress-energy tensor 
and a finite Euclidean action by adding an appropriate surface term to the 
bulk action, although the solution is not asymptotically AdS. For the 
solution (\ref{2eq6}), we find that the suitable counterterm is 
\begin{equation}
\label{2eq8}
S_{\rm ct}=-\frac{1}{8\pi G_{n+2}}\int d^{n+1}x\sqrt{-h} \frac{c_0}
 {l_{\rm eff}},
\end{equation}
where
\begin{equation}
\label{2eq9}
c_0= n\sqrt{\frac{N(n+1)}{N(n+2)-1}}, \ \ \ \frac{1}{l_{\rm eff}}
   = \sqrt{\frac{V_0e^{-a\phi}} {n(n+1)}}.
\end{equation}
With this counterterm, the quasilocal stress-energy tensor of gravitational
field  at the boundary $r$ with an induced metric $h$ is 
\begin{equation}
\label{2eq10}
T_{ab}=\frac{1}{8\pi G_{n+2}}\left ( K_{ab}-Kh_{ab} -\frac{c_0}{l_{\rm eff}}
      h_{ab} \right),
\end{equation}
where $K$ is the extrinsic curvature of the boundary $h$. Calculating the
tensor yields
\begin{eqnarray}
\label{2eq11}
&& 8\pi G_{n+2}T_{tt}= \frac{nN}{2}\left( \frac{V_0 e^{a\phi_0}}{nN(N(n+2)-1)}
       \right)^{1/2} \frac{m}{\sqrt{2nN(1-N)}} \frac{1}{r^{(n-1)N}}
   +\cdots, \nonumber \\
&& 8\pi G_{n+2}T_{ij}=\delta_{ij} \frac{2N-1}{2}\left(\frac{V_0e^{-a\phi_0}}
    {nN(N(n+2)-1)}\right)^{-1/2}\frac{m}{\sqrt{2nN(1-N)}}\frac{1}{r^{(n-1)N}}
  +\cdots,
\end{eqnarray} 
where the ellipses denote higher order terms, which have  no contributions
when we move the boundary to the spatial infinity ($r\to \infty$).

Given a well-defined quasilocal stress-energy tensor of gravitational field,
one can calculate some conserved charges associated with Killing 
vectors~\cite{Bala}. The energy of gravitational field is a conserved charge
 associated with a timelike Killing vector. Applying this to the
solution (\ref{2eq6}), we obtain the mass $M$ of the domain-wall black holes  
\begin{equation}
\label{2eq12}
M \equiv  \int_{r\to \infty} d^nx r^{nN}f^{-1/2}T_{tt}=
   \frac{nN}{\sqrt{2nN(1-N)}}\frac{mV_n}{16\pi G_{n+2}}.
\end{equation}
where $V_n$ denotes the volume of the domain wall, namely, the volume of
line element $dx_n^2$ in the solution (\ref{2eq6}). We see that as 
we explained above, the integration constant $m$ is indeed related to 
the mass of the domain-wall black holes.

Next we calculate the stress-energy tensor of the thermal QFT corresponding 
to the domain-wall black hole  solution (\ref{2eq6}). The boundary metric 
$\gamma_{ab}$ of the spacetime, in which the boundary QFT resides, can be 
determined as follows,  
\begin{equation}
\label{2eq13} 
\gamma _{ab} =\lim _{r\to \infty}\frac{1}{r^{2N}}h_{ab}
   = -\frac{V_0e^{-a\phi_0}}{nN(N(n+2)-1)} dt^2 +dx_n^2.
\end{equation}
Rescaling the time coordinate $t$, one has 
\begin{equation}
\label{2eq14}
\gamma_{ab}= -d\tau^2 +dx_n^2.
\end{equation}
The boundary stress-energy tensor $\tau_{ab}$ of the QFT can be achieved
through the relation \cite{Myers} 
\begin{equation}
\label{2eq15}
\sqrt{-\gamma}\gamma^{ab}\tau_{bc}=\lim_{r \to \infty}\sqrt{-h}h^{ab}T_{bc}.
\end{equation}
Substituting (\ref{2eq11}) into the above formula, we finally arrive at
\begin{eqnarray}
\label{2eq16}
&& \tau_{\tau\tau}=\frac{M}{cV_n},\ \ \ \
   c=\left(\frac{V_0e^{-a\phi_0}} {nN(N(n+2)-1)}\right)^{1/2}, 
 \nonumber \\
&& \tau_{ij}=\delta_{ij} \frac{M(2N-1)}{cnNV_n}. 
\end{eqnarray}
This can be explained as the vacuum expectation value of the stress-energy
tensor of the QFT residing in the spacetime (\ref{2eq14}). 
From (\ref{2eq16}) we obtain the equation of state of the QFT
\begin{equation}
\label{2eq17}
p =\frac{2N-1}{nN}\rho,
\end{equation}
relating the pressure $p$ and the energy density $\rho$ of the QFT.
We can see clearly from (\ref{2eq17}) that when $N=1$, one has $p=\rho/n$,
an equation of state for CFTs. This is consistent with the fact that
 when $N=1$, the solution (\ref{2eq6}) describes an $(n+2)$-dimensional
 AdS Schwarzschild black
 hole, to which there is a $(n+1)$-dimensional  CFT corresponding. 
When $N=1/2$, the pressure $p$ vanishes.
The D5-branes and NS5-branes are just this  case \cite{CO}.
When $N<1/2$, the pressure becomes negative and in  this case, the 
domain-wall/QFT correspondence might be invalid. The D$p$-branes with $p>5$
belong to this case \cite{CO}, for which we know gravity does not decouple
in the usual decoupling limit.     
 
Before closing this section, let us calculate the Hawking temperature 
$T_{\rm HK}$ and the entropy $S$ of the domain-wall black holes (\ref{2eq6}).
A simple calculation yields 
\begin{eqnarray}
\label{2eq18}
&& T_{\rm HK} = \frac{N(n+2)-1}{4\pi} c^2 r_+^{2N-1}, \nonumber \\
&& S= \frac{r_+^{nN}}{4G_{n+2}}V_n,
\end{eqnarray}
where $c^2$ is  the one in (\ref{2eq16}) and $r_+$ is the horizon radius of 
the black hole. It is easy to check that these
quantities  satisfy, $dM =T_{\rm HK} dS$, the first law of black hole 
thermodynamics.


\sect{Brane cosmology in the domain-wall background}

In this section following \cite{Savo,Gubser}\footnote{The motion of a 
brane (domain wall) in the AdS black hole background has also been 
discussed in \cite{Krau,Ida}, but in those papers the brane is embedded to the
black hole background, and does not act as the edge of the bulk.}, we discuss 
the dynamics of a brane universe in the background of the dilaton domain-wall
black holes (\ref{2eq6}). The dynamics of the ($n+1$)-dimensional brane is 
governed by the action
\begin{equation}
\label{3eq1}
S_b =-\frac{1}{8\pi G_{n+2}}\int d^{n+1}x\sqrt{-h}K 
           -\int d^{n+1}x \sqrt{-h}\sigma,
\end{equation}
where the first term is the Gibbon-Hawking boundary term and $\sigma $ is the
tension of the brane. Viewing the brane as the edge at a finite $r$ of the 
bulk (\ref{2eq6}), one has  the equation of motion of the brane    
\begin{equation}
\label{3eq2}
K_{ab}= \frac{8\pi G_{n+2}}{n}\sigma h_{ab}.
\end{equation}
Furthermore, in order for the variation principle to be well-defined for the 
total action, which is sum of the bulk 
action (\ref{2eq1}) and the brane action (\ref{3eq1}), the tension $\sigma$
of the brane must depend on the dilaton and satisfies \cite{Chamb}
\begin{equation}
\label{3eq3}
n^a\partial_a \phi = 16\pi G_{n+2} \frac{\partial \sigma}{\partial \phi},
\end{equation}
where $n^a$ is a unit  normal vector  to the brane.

Now let us specify the location of the brane as $r=r(t)$. If we introduce a
cosmic time $\tau$ so that $t=t(\tau)$ and $r=r(\tau)$ and satisfies
\begin{equation}
\label{3eq4}
f(r)\left (\frac{dt}{d\tau}\right)^2 -\frac{1}{f(r)}\left(\frac{dr}
 {d\tau}\right)^2 =1,
\end{equation}
from Eq.~(\ref{2eq6}), one can see that the induced metric on the brane 
then takes the form
\begin{equation}
\label{3eq5}
ds^2 =-d\tau^2 +R^2(\tau)dx_n^2. 
\end{equation}
Note that here $R$ is a function of $\tau$ through the relation $R=r^N$.
The equation (\ref{3eq5}) is a standard form of the metric for an 
($n+1$)-dimensional flat ($k=0$) FRW universe.

From Eqs.~(\ref{3eq2}) and (\ref{3eq4}) we obtain the equation of motion
of  the scale factor $R$  
\begin{equation}
\label{3eq6}
H^2 =\left(\frac{8\pi G_{n+2}\sigma}{n}\right)^2 -\frac{N^2}{R^{2/N}}f(R),
\end{equation}
where $H=\dot R/R$ is the Hubble constant and the overdot stands for 
derivative with respect to the cosmic time $\tau$.  We find from 
Eqs.~(\ref{3eq2}) and (\ref{3eq3}) that the tension $\sigma$ is a function 
of $\phi$ as (see also \cite{Chamb})
\begin{equation}
\label{3eq7}
\sigma =\sigma_0 e^{-a\phi/2},
\end{equation}
where $\sigma_0$ is a constant. Substituting the tension into 
Eq.~(\ref{3eq6}) yields
\begin{equation}
\label{3eq8}
H^2 = \left(\frac{8\pi G_{n+2}\sigma_0}{n}\right)^2 e^{-a\phi_0}R^{2-2/N}
       -\frac{V_0N^2 e^{-a\phi_0}R^{2-2/N}}{nN(N(n+2)-1)}
      +\frac{mN^2}{\sqrt{2nN(1-N)}R^{n+1/N}}.
\end{equation}
Now we choose the constant $\sigma_0$ so that the first two terms in 
Eq.~(\ref{3eq8}) cancel with each other. That is, we take 
\begin{equation}
\label{3eq9}
\sigma_0 =\frac{1}{8\pi G_{n+2}}\sqrt{\frac{nN V_0}{N(n+2)-1}}.
\end{equation}
This choice corresponds to the fine-tuning of the brane 
 in the case of AdS Schwarzschild black holes \cite{Savo,Gubser}. In 
 such a choice as (\ref{3eq9}), we note that the brane tension term in 
(\ref{3eq1}) is just equal to the surface counterterm (\ref{2eq8}). 
The surface
counterterm (\ref{2eq8}) makes the quasilocal stress-energy tensor of 
gravitational field (\ref{2eq10}) be well-defined, while here the  
effective cosmological constant on the brane vanishes under the choice
(\ref{3eq9}).

Given this choice, the Friedmann equation (\ref{3eq8}) then reduces to
\begin{equation}
\label{3eq10}
H^2 = \frac{16\pi G_{n+2} MN}{nV_n}\frac{1}{R^{n+1/N}},
\end{equation} 
where the integration constant $m$ has been replaced by the mass $M$ of the
domain-wall black hole. Further, differentiating the equation (\ref{3eq10})
with respect to the cosmic time $\tau$, one can have the equation of the
time derivative of $H$,
\begin{equation}
\label{3eq11}
\dot H = -\left(n +\frac{1}{N}\right) \frac{8\pi G_{n+2} MN}{nV_n}
     \frac{1}{R^{n+1/N}}.
\end{equation}
The evolution of the brane universe is easily determined by integrating the 
equation (\ref{3eq10}), which gives
\begin{equation}
\label{3eq12}
R=\left( \frac{nN+1}{2N}\sqrt{\frac{16\pi G_{n+2}MN}{nV_n}}\tau 
     \right)^{2N/(nN+1)}.
\end{equation}
Here an integration constant has been set to zero, for one can always 
do so by shifting the cosmic time $\tau$. The solution (\ref{3eq12})
describes an expanding flat ($k=0$) universe.

We now turn to studying  the holographic properties associated with the
 brane universe. Note that the mass $M$ of the domain-wall black holes,
namely, the energy of thermal QFT, is measured with respect to the coordinate 
time $t$ in the metric (\ref{2eq13}).  Measured in the  
coordinates (\ref{3eq5}),  the energy of the QFT is 
\begin{equation}
\label{3eq13}
E =\frac{1}{cR}M,
\end{equation}
where $c$ is given by Eq.~(\ref{2eq16}). Let us further note that the
gravitational constant $G_{n+2}$ is the one in ($n+2$) dimensions. That is, 
it is the bulk gravitational constant. We find that the gravitational constant
$G_{n+1}$ on the brane is 
\begin{equation}
\label{3eq14}
G_{n+1}=\frac{(n-1)Nc}{R^{(1-N)/N}}G_{n+2},
\end{equation}
which has been obtained by adding a small amount of stress energy on the 
brane and then comparing the equation of motion of the brane with the 
standard Friedmann equation in ($n+1$) dimensions\footnote{For the equation 
of motion of brane with localized matter see \cite{Krau,Ida}.}. It is 
interesting to note that the brane gravitational constant
depends on the scale factor $R$. It implies that one can view the brane 
universe in the background of the dilaton domain-wall black holes as a universe
model of varying gravitational constant. As a consistency check, it is easy 
to see that when $N=1$, {\it i.e.}, the bulk geometry is the AdS 
Schwarzschild  solution,  the brane gravitational constant $G_{n+1}$ becomes a 
true constant and the relation $G_{n+1}=(n-1)G_{n+2}/L$ can be 
recovered~\cite{Gubser,Savo}.

Using the gravitational constant $G_{n+1}$ and energy $E$ on the brane, 
we find that the equation (\ref{3eq10}) can be rewritten as 
\begin{equation}
\label{3eq15}
H^2 = \frac{16\pi G_{n+1}}{n(n-1)}\frac{E}{V},
\end{equation}
a standard form of the Friedmann equation for an ($n+1$)-dimensional 
flat ($k=0$) universe, where $V=R^nV_n$ is the volume of the universe. 
So we see that the energy of brane universe is provided by the thermal 
QFT corresponding to the domain-wall black holes. For the time derivative 
of the Hubble constant $H$, we have
\begin{equation}
\label{3eq16}
\dot{H} = -\frac{(nN+1)}{nN}\frac{8\pi G_{n+1}}{(n-1)} \frac{E}{V}. 
\end{equation}
Comparing  this with its standard form, $\dot {H} =-
  \frac{8\pi G_{n+1}}{(n-1)}\left(\frac{E}{V}+p\right)$, we obtain the 
equation of state of matter filling  the brane universe
\begin{equation}
\label{3eq17}
p = \frac{1}{nN}\rho,  \ \ \ \rho =\frac{E}{V}.
\end{equation}
Comparing  this with the equation (\ref{2eq17}) of state of the QFT, obtained 
in the previous section, one can see that they can be matched only 
when $N=1$, namely, the bulk is the AdS Schwarzschild 
black holes (recall that in this case the gravitational constant on the
brane is a true constant). This result seems strange, since 
naively one might expect that 
the equation of motion of the brane should be the Friedmann equation with the 
equation (\ref{2eq17}) of state of the QFT. This mismatch  might be   
caused by the varying gravitational constant on the brane when $N\ne 1$, 
since we derived  Eq.~(\ref{3eq17}) from Eqs. (\ref{3eq15}) and (\ref{3eq16})
 by comparing them with corresponding equations in the Einstein gravity. 
 The fact, however, is that the gravity on the brane is a 
Brans-Dicke-like gravity theory: The gravitational constant is a dynamical 
one. Therefore we should expect that the dynamical gravitational constant 
makes some contributions to the pressure and energy density on the brane. 
This might be one of reasons why there is a difference between (\ref{3eq17})
and (\ref{2eq17}). Another possible cause is that there  might 
be some contribution of the dilaton field. 
In  Ref.~\cite{Maeda}, the gravitational equations on the brane
have been given when the bulk action includes a dilaton potential. In our 
bulk solution (\ref{2eq6}), the dilaton field $\phi$ is static in the 
coordinates (\ref{2eq6}), but is not static in the coordinates on the brane
(\ref{3eq5}) through the relation $r=r(\tau)$. From the gravitational equations 
in \cite{Maeda} one can see that the dynamical dilaton field will also make 
 contributions to the energy density and pressure on the brane. Obviously,
it is quite necessary  to further clarify  the origin of the difference between 
Eq.~(\ref{2eq17}) and Eq.~(\ref{3eq17}), which is currently under 
investigation.

Following Ref.~\cite{Verl}, we define the Hubble 
entropy bound $S_{\rm H}$, the Bekenstein-Verlinde entropy bound $S_{\rm BV}$,
and the Bekenstein-Hawking  entropy bound $S_{\rm BH}$, 
\begin{equation}
\label{3eq18}
S_{\rm H}=(n-1)\frac{HV}{4G_{n+1}}, \ \ S_{\rm BV} =\frac{2\pi}{n}ER, \ \
S_{\rm BH}=(n-1)\frac{V}{4G_{n+1}R}. 
\end{equation}
The Friedmann equation (\ref{3eq15}) can then be expressed  as\footnote{For
a closed  FRW universe, the corresponding form is 
$S_{\rm H}^2=S_{\rm BH}(2S_{\rm BV}- S_{\rm BH})$. See \cite{Verl}.}
\begin{equation}
\label{3eq19}
S_{\rm H}^2 =2S_{\rm BH} S_{\rm BV}.
\end{equation}
If one uses  the Hubble entropy density, $s_{\rm H}=S_{\rm H}/V$,  and
the energy density $\rho$, the
Friedmann equation (\ref{3eq15}) can be further rewritten as 
\begin{equation}
\label{3eq20}
s_{\rm H}^2 =\left(\frac{4\pi}{n}\right)^2 \frac{n(n-1)}{16\pi G_{n+1}}\rho.
\end{equation} 
We now consider a special moment when the brane crosses the black hole 
horizon of the background (\ref{2eq6}), {\it i.e.}, $R=r_+^n$.
 it is easy to show that the Hubble 
entropy bound is just saturated by the entropy of the domain-wall black hole,
\begin{equation}
\label{3eq21}
S_{\rm H} =(n-1)\frac{HV}{4G_{n+1}}=\frac{V_nr_+^{nN}}{4G_{n+2}}=S,
   \ \ {\rm at}\ \  R=r_+^N, 
\end{equation}
as the case of AdS Schwarzschild black holes.
The temperature on the brane (that is, the temperature of brane universe) is 
\begin{equation}
\label{3eq22}
T =\frac{1}{cR}T_{\rm HK}=\frac{N(n+2)-1}{4\pi}c r_+^{N-1}, \ \ {\rm at}\ \
  R=r_+^N.
\end{equation}
As in \cite{Verl}, if we define the limiting temperature $T_{\rm H}$ in 
terms of the derivative of the Hubble constant, $T_{\rm H} \equiv 
-\dot{H}/(2\pi H)$, it is then
\begin{equation}
\label{3eq23}
T_{\rm H} = \frac{nN+1}{4\pi}c r_+^{N-1}, 
 \ \  {\rm at} \ \ R=r_+^N. 
\end{equation}
Comparing (\ref{3eq23}) with (\ref{3eq22}), one can see that
 when the brane crosses the horizon of the domain-wall black holes,
the two temperatures $T$ and $T_{\rm H}$ are equal only if $N=1$. In other 
words, that the limiting temperature $T_{\rm H}$ is saturated by the 
temperature of the domain-wall black holes holds only when the bulk is the 
AdS Schwarzschild geometry \cite{Savo}. This can be related to the fact that 
one has a varying gravitational constant on the brane  when $N\ne 1$.

The total entropy of the brane universe is the entropy of the domain-wall
black hole, which is a constant during the evolution of the universe. But the
entropy density is not a constant, which is
\begin{equation}
\label{3eq24}
s =\frac{r_+^{nN}}{4G_{n+2}R^n},
\end{equation}
and the energy density of the brane universe is
\begin{equation}
\label{3eq25}
\rho \equiv \frac{E}{V} =\frac{cnN r_+^{(n+2)N-1}}{16\pi G_{n+2} R^{n+1}}.
\end{equation}
We find that the entropy density can be related to its energy 
density as
\begin{equation}
\label{3eq26}
s^2 = \left (\frac{4\pi}{n}\right)^2 \gamma \rho,\ \ \ \ 
\gamma = \frac{n(n-1)}{16\pi G_{n+1}}\frac{r_+^{(n-2)N+1}}{R^{n-2 +1/N}},
\end{equation}
which can be viewed as a deformed form of the Cardy-Verlinde formula 
of CFTs\footnote{The corresponding Cardy-Verlinde formula of CFTs in the 
closed FRW universe is $s^2=\left(\frac{4\pi}{n}\right)^2 \gamma 
\left(\rho -\frac{\gamma}{R^2}\right)$. Here the term $\gamma/R^2$ is the
contribution of Casimir effect.  See \cite{Savo}.}: Since we are 
considering a flat universe, there is no contribution of
Casimir effect.
This relation (\ref{3eq26}) always holds during the evolution of the brane 
universe. But at the moment when the brane crosses the background horizon, 
we note that this equation (\ref{3eq26}) coincides with the Friedmann 
equation (\ref{3eq20}), as the case of the AdS Schwarzschild black
holes \cite{Savo}. Thus the same can be stated here: The Friedmann equation 
(\ref{3eq20})  somehow knows the entropy formula of QFTs filling the brane
universe.

Next we make some remarks about  the result (\ref{3eq26}). 
The Cardy formula \cite{Cardy} is supposed to hold only for 
$(1+1)$-dimensional CFTs.  After some appropriate identifications, however, 
Verlinde~\cite{Verl} argued that the Cardy formula  also holds for CFTs
in arbitrary dimensions, resulting in the so-called Cardy-Verlinde formula.
The Cardy-Verlinde formula has been checked to be valid  for
CFTs with AdS duals (for example, see \cite{Verl},\cite{Klem},\cite{Cai1}
and \cite{Birm1}). Here we found that a Cardy-Verlinde-like formula 
(\ref{3eq26}) holds
for a QFT corresponding to the domain-wall black holes (\ref{2eq6}).    
At first sight, this result seems strange since one does not expect the 
thermodynamics of QFTs obeys a Cardy-Verlinde-like formula.  At this point,
it seems helpful to recall the recent claim that  
the entropy of asymptotically flat black holes can also be expressed by 
a modified Cardy formula \cite{Klem,Youm}, and  in particular the Carlip's 
result that for black holes in any dimension the Bekenstein-Hawking
entropy can be reproduced using the Cardy formula~\cite{Carlip}.  Carlip
obtained his result by considering general relativity on a manifold with 
boundary. He found that the constraint algebra of general relativity may 
acquire a central extension, which can be calculated using covariant phase
space techniques. When the boundary is a (local) Killing horizon, a natural
set of boundary conditions leads to a Virasoro subalgebra with a calculable
central charge. He then used conformal field theory methods to determine
the density of states at the boundary, which yields the expected entropy
of black holes. Obviously, the Carlip's method is quite different from 
the Verlinde's argument. And the Carlip's method is applicable for a  
wider class of black holes.  But both methods lead to the same conclusion 
that the entropy of black holes can be expressed by a modified Cardy formula.
These results strongly imply that the thermodynamics of  black 
holes is of some common features of thermodynamics of $(1+1)$-dimensional 
CFTs \cite{Klem}. Therefore, our result (\ref{3eq26}) looks strange from the 
point of view of thermodynamics of QFTs, but does not from the point of view of
black hole thermodynamics and is  consistent with the argument 
advocated by Carlip that the entropy of all black holes can be expressed in
terms of a modified Cardy formula~\cite{Carlip}.

\sect{Conclusions}

We have found a class of domain-wall black hole solutions (AdS Schwarzschild 
black hole is included as a special case) in a  dilaton gravity with a 
Liouville-type dilaton potential. Using the method of surface counterterm  
we have obtained the stress-energy tensor of the boundary QFT corresponding 
the domain-wall black holes in the domain-wall/QFT correspondence. The entropy
and Hawking temperature of the domain-wall black holes  have been also 
calculated.  

We have considered a brane universe in the background of the 
domain-wall black holes. With the choice that the tension term of the bane is
just the surface counterterm, which is introduced in order to acquire 
the finite boundary stress-energy tensor of the QFT, we have found that 
the equation of motion of the brane
can be mapped to a standard form of the Friedmann equation for a flat ($k=0$)
FRW universe. The gravitational constant on the brane depends on the scale 
factor of the brane universe. It means that the brane universe in the 
background of dilaton domain-wall black holes can be viewed as a universe
model with a varying gravitational constant. By comparing the FRW
equations of brane universe with the standard FRW equations in Einstein 
gravity,  
we have found that the resulting equation of state of matter filling the 
brane universe cannot be matched  to  the one of the QFT obtained from the
surface counterterm approach unless $N=1$. That is, they can be identified 
with each other only when the bulk is the AdS Schwarzschild black 
hole (recall in this case that the gravitational constant on the brane is a
true constant). This might be caused by the varying gravitational constant on 
the brane. Because the gravity on the brane in fact is a Brans-Dicke-like 
gravity theory, the dynamical gravitational constant will also make some 
contributions to the pressure and energy density  on the brane. The 
contribution of the dilaton field might be also one of reasons  of the
mismatch.  Because of this mismatch, the limiting temperature
expressed in terms of the Hubble constant and its time derivative is not 
equal to the Hawking temperature of the domain-wall black holes when the
brane crosses the horizon of black holes.  However, 
a  Cardy-Verlinde-like formula (\ref{3eq26}) has been found, which relates the
entropy density of the QFT to its energy density. At the moment when
the brane crosses the black hole horizon of the background,  the  
Cardy-Verlinde-like formula coincides with the Friedmann equation of the 
brane, and the Hubble entropy bound is still saturated by the entropy of
the domain-wall black hole. These are the  same as those in the case of AdS 
Schwarzschild black hole.

Note that the Cardy-Verlinde formula can express the 
entropy of any dimensional black holes in asymptotically AdS spacetimes.
It has been claimed  recently in\cite{Klem,Youm} that the entropy of 
asymptotically flat black holes can also be expressed in terms of a 
modified Cardy formula. We have found here that the Cardy-Verlinde-like 
formula (\ref{3eq26}) can describe the entropy of a class of dilaton 
domain-wall black holes (\ref{2eq6}), which is neither asymptotically AdS, 
nor asymptotically flat. These results altogether support the argument 
advocated by Carlip that the entropy of all black holes can be expressed in 
terms of a modified Cardy formula.
It seemingly  implies that the thermodynamics of black holes has a close 
connection with that of $(1+1)$-dimensional CFTs.

\section*{Acknowledgments}

We would like to thank Prof. N. Ohta for a critical reading of manuscript. 
The work of RGC  was supported in part by a grant from Chinese Academy of 
Sciences, and in part by the Japan Society for the
Promotion of Science and  Grants-in-Aid for Scientific Research
Nos. 99020 and 12640270. YZZ was supported in part by National Natural 
Science Foundation of China under Grant No. 10047004, and by Ministry of
Science and Technology of Modern Grant No. NKBRSF G19990754.


\end{document}